%% file: sample-sigconf.tex
\documentclass[sigconf]{acmart}

\usepackage{tikz}
\usepackage{pgfplotstable}
\usepackage{pgfplots}
\usetikzlibrary{intersections}
\usepackage{subcaption}

\AtBeginDocument{%
  \providecommand\BibTeX{{%
    \normalfont B\kern-0.5em{\scshape i\kern-0.25em b}\kern-0.8em\TeX}}}

\setcopyright{acmcopyright}
\copyrightyear{2023}
\acmYear{2023}
\acmDOI{XXXXXXX.XXXXXXX}

\acmConference[MMAsia '23]{ACM Multimedia Asia}{December 06--08, 2023}{Tainan, Taiwan}
\acmBooktitle{ACM Multimedia Asia (MMAsia ’23), December 06--08, 2023, Tainan, Taiwan}
\acmPrice{15.00}
\acmISBN{978-1-4503-XXXX-X/18/06}

\acmSubmissionID{4}

\begin{document}

\title{Audio Time-Scale Modification with Temporal Compressing Networks}

\author{Ernie Chu}
\authornote{Part of the work was done while Ernie and Ju-Ting were students at National Sun Yat-sen University.}
\affiliation{%
  \institution{Research Center for Information Technology Innovation,\\Academia Sinica}
  \city{Taipei}
  \country{Taiwan}}
\email{shchu@citi.sinica.edu.tw}

\author{Ju-Ting Cheng}
\authornotemark[1]
\affiliation{%
  \institution{National Cheng Kung University}
  \city{Tainan}
  \country{Taiwan}}

\author{Chia-Ping Chen}
\affiliation{%
  \institution{National Sun Yat-sen University}
  \city{Kaohsiung}
  \country{Taiwan}}
\email{cpchen@cse.nsysu.edu.tw}


\begin{abstract}
\input{sections/0-abs}
\end{abstract}





\maketitle

\section{Introduction}
\input{sections/1-intro}

\section{Related Work}
\input{sections/2-related-work}

\section{Method}
\input{sections/3-method}

\section{Experiment}
\input{sections/4-exp}

\section{Conclusion and limitation}
\input{sections/5-conclusion}

\begin{acks}
This work is supported by National Science and Technology Council (formerly Ministry of Science and Technology), Taiwan (R.O.C), under Grants no. 110-2813-C-110-050-E.
\end{acks}

\bibliographystyle{ACM-Reference-Format}
\bibliography{sample-tsm}










\end{document}

%% file: sections/0-abs.tex
We propose a novel approach for time-scale modification of audio signals. Unlike traditional methods that rely on the framing technique or the short-time Fourier transform to preserve the frequency during temporal stretching, our neural network model encodes the raw audio into a high-level latent representation, dubbed Neuralgram, where each vector represents 1024 audio sample points. Due to a sufficient compression ratio, we are able to apply arbitrary spatial interpolation of the Neuralgram to perform temporal stretching. Finally, a learned neural decoder synthesizes the time-scaled audio samples based on the stretched Neuralgram representation. Both the encoder and decoder are trained with latent regression losses and adversarial losses in order to obtain high-fidelity audio samples. Despite its simplicity, our method has comparable performance compared to the existing baselines and opens a new possibility in  research into modern time-scale modification. Audio samples can be found on our website\footnote{\url{https://tsmnet-mmasia23.github.io}.}. 

%% file: sections/1-intro.tex
\begin{figure}[t]
  \centering
  \includegraphics[width=.95\linewidth]{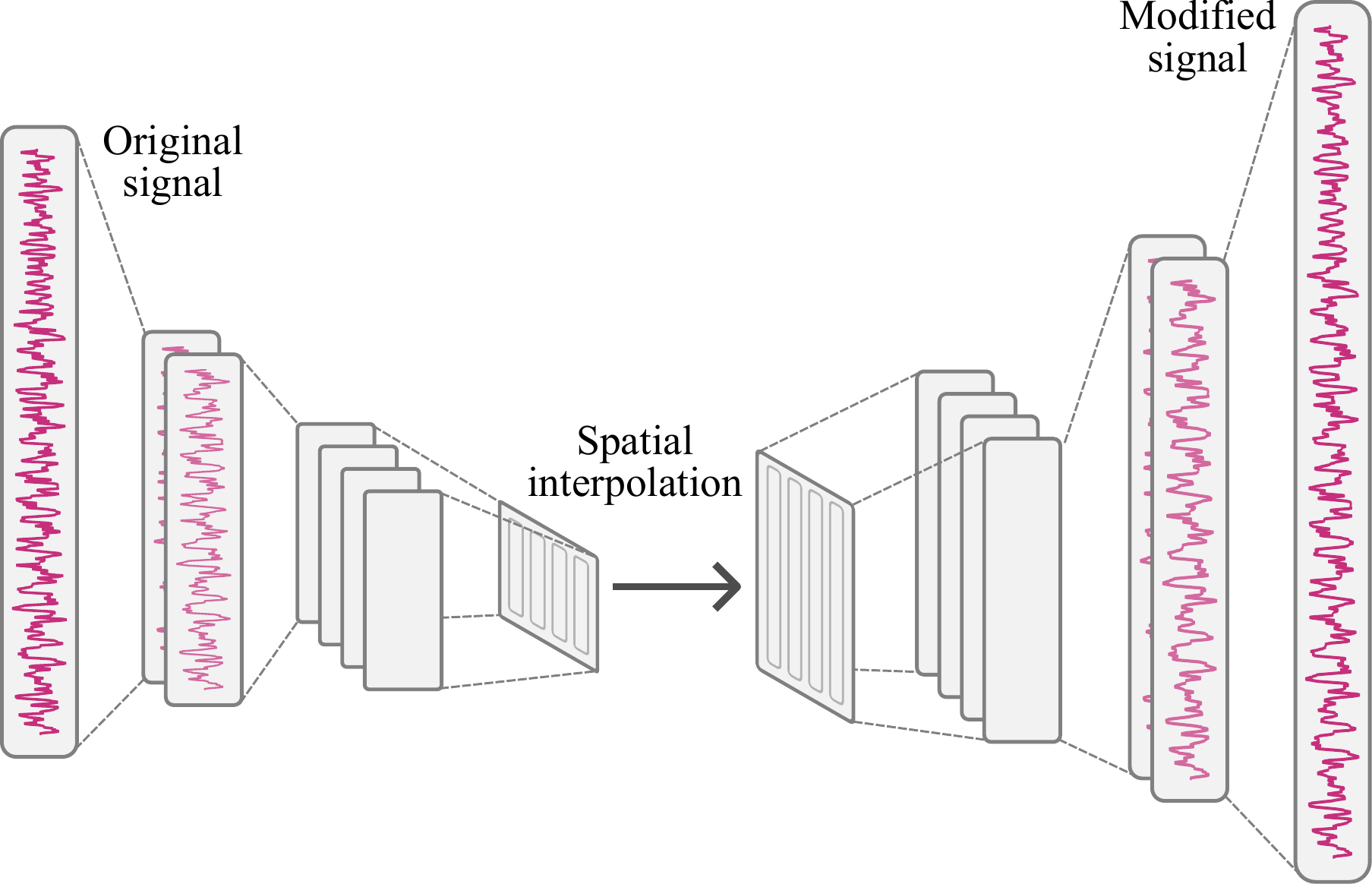}
  \caption{Overall architecture for the proposed method. TSM-Net temporally stretches an audio signal in a simple approach.}
  \label{fig:arch}
\end{figure}

With the advancement of technologies and digitalization, we can store and reproduce multimedia content. We can even manipulate materials in ways that we couldn't imagine before. For example, image resizing and video editing change the dimensions of digital pictures spatially and temporally, respectively. Another ubiquitous application regarding audio signals is time-scale modification (TSM), which is used in our daily life. It is also known as playback speed control in video streaming platforms, such as YouTube.

With the power of artificial intelligence (AI) and modern computation hardware, pragmatic AI tools are emerging in multimedia domains, such as image super-resolution \cite{led17} and motion estimation, motion compensation \cite{bao21}, etc. However, as far as we know, methods leveraging AI to refine the TSM algorithm and improve the quality of the synthetic audio, have not been well-studied.

Naive resampling of the raw audio signal to get scaled versions causes serious pitch-shifting effects because the wavelength of each frequency component scales proportionally with the duration of the overall sample. TSM has been a research topic aiming to alleviate the pitch-shifting effects when stretching audio samples. Several methods have been proposed, including time-domain approaches and spectral approaches. The ultimate goal of TSM is to synthesize high-quality audio that is perceptually indistinguishable from real-world recordings. Despite the efforts of early research and wide applications in the market, existing methods merely achieve this goal for some speech-only audio. When it comes to more complicated content like music, users face a trade-off in quality between the harmonic source and percussive source. Our aim is to develop a universal method that can accommodate all types of audio content while minimizing the occurrence of artifacts to the greatest extent possible.

In our work, we use an architecture similar to the autoencoder \cite{kra91}, which encodes the data into high-level (and typically low-dimensional) latent vectors and faithfully reconstructs the original data. In our model, the dimension of the latent vectors is 1024 times smaller than the original one, which means one sample in the latent vector can represent more than an entire wave in the raw audio waveform. Since the smallest unit encompasses more than one wave, we can apply arbitrary spatial interpolation on the latent vector to stretch the audio, without worrying about the changes in frequency components and the pitches. Finally, we decode the resized latent vector to obtain the time-scaled audio waveform. Our overall architecture is illustrated in Figure \ref{fig:arch}. To synthesize high-fidelity audio samples, we use the adversarial losses to train our autoencoder \cite{goo14}. Multi-scale discriminative networks are employed to distinguish between real and generated data. We will delve into the details of the training in Section \ref{sec:method}.

In summary, our main contribution is to propose a simple yet powerful data-driven method that shows comparable performance on various kinds of audio contents. We would also like to reignite the research on TSM in the Machine Learning era. Through the improvement of this fundamental tool in audio processing, we believe a wide range of applications can directly benefit.

%% file: sections/2-related-work.tex
In this section, we comprehensively review the existing approaches for TSM, both in the time domain and the spectral domain. Additionally, we review the fundamental tool in audio generation, the neural vocoder. This tool would be an essential component for the proposed method.

\subsection{Time-domain approach}
\label{sec:td-tsm}
\begin{figure}[t]
  \centering
  \includegraphics[width=.4\textwidth]{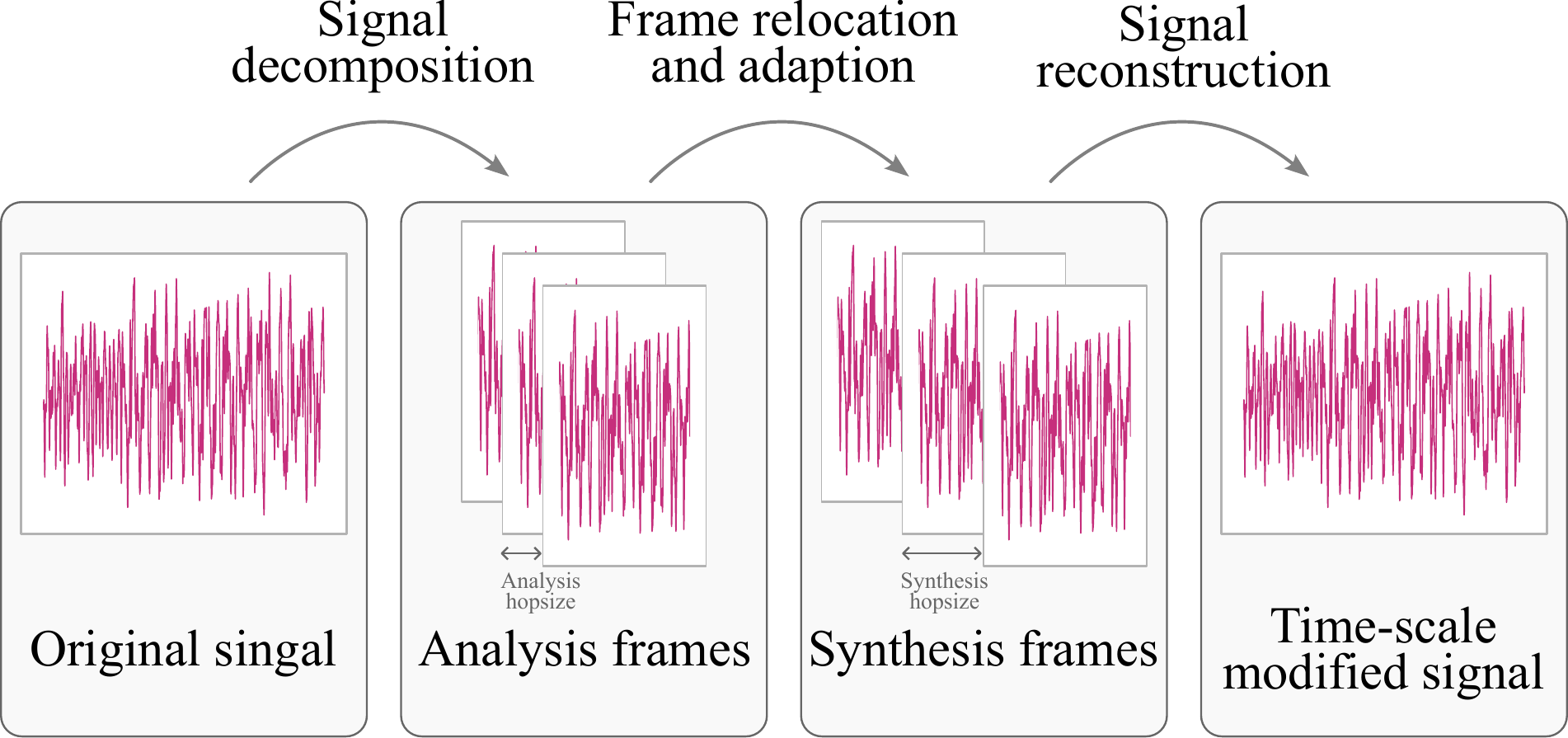}
  \caption{Generic processing pipeline of time-domain time-scale modification 
(TSM) procedures.}
  \label{fig:tsm-pipeline}
\end{figure}

The main idea of TSM is that instead of scaling the raw waveforms on the time axis, which leads to pitch shifts due to the changes in wavelengths, we segment the audio into small chunks of fixed length, also known as frames or windows, in order to preserve the wavelength. The adjacent frames are overlapped and rearranged to minimize boundary breakage after processing, as shown in Figure \ref{fig:tsm-pipeline}.

\begin{figure}[t]
  \centering
  \vspace{0.5cm}
  \includegraphics[width=0.4\textwidth]{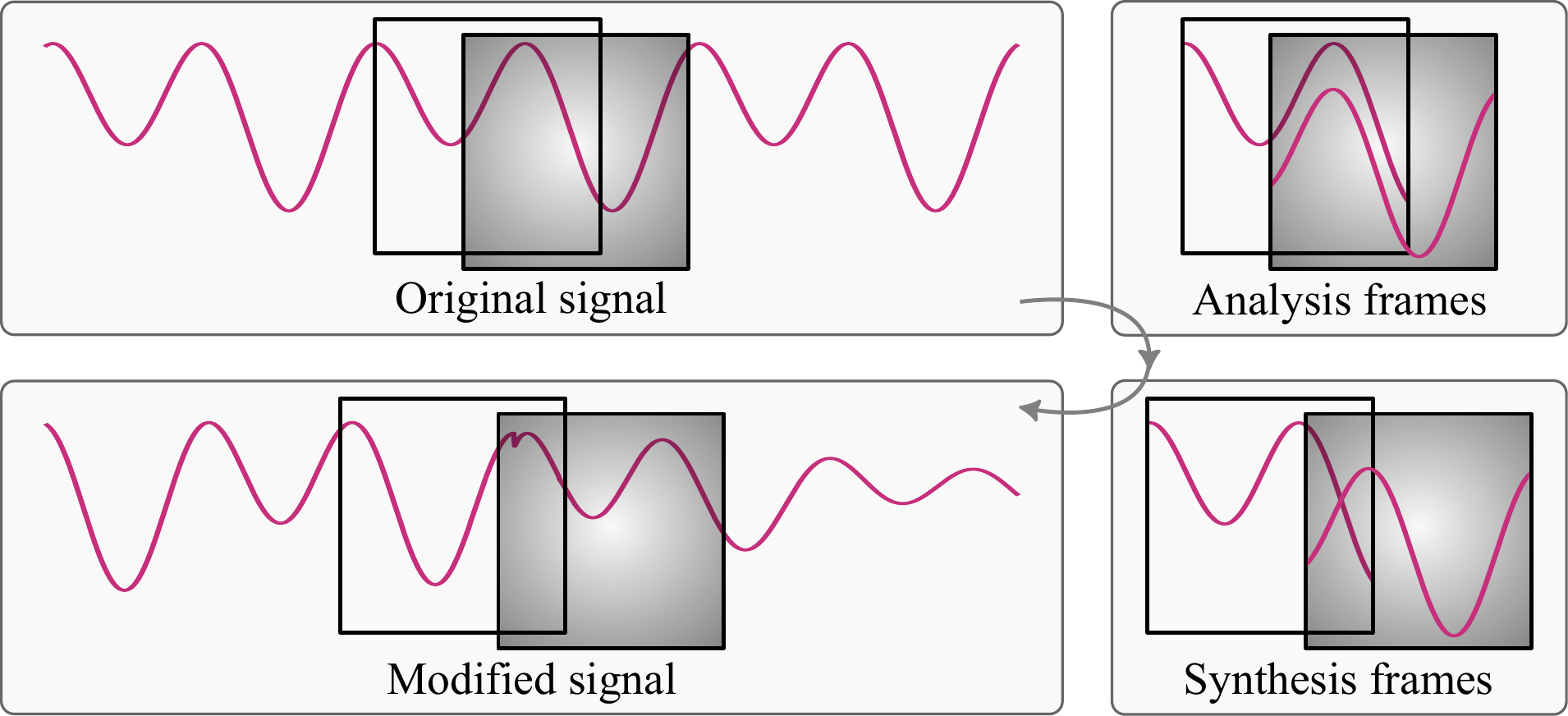}
  \caption{An illustration of the harmonic alignment problem. The black boxes demonstrate the rearrangement of the frames. An unconstrained scale ratio would lead to serious interference.}
  \label{fig:harmonic-alignment}
\end{figure}

The original distance between the start of each frame is called the analysis hop size. Once the frames are relocated, this distance becomes the synthesis hop size, and the ratio of the analysis hop size to the synthesis hop size determines the speed at which the audio is accelerated or decelerated \cite{dri16}. Additionally, the Hann window \cite{ess86} is usually applied to each analysis frame to maintain the amplitude of the overlapped areas. One of the main challenges in this approach is the harmonic alignment problem, which is illustrated in Figure \ref{fig:harmonic-alignment}.

When significant periodicities are present, an unconstrained ratio of the analysis to synthesis hop size can cause a discrepancy with the original waveform. Specifically, the phases of the frequency components within the frames do not synchronize properly, resulting in significant interference. Several solutions have been proposed to address the synchronization problem \cite{hej91,eri90,ver93}. However, the resulting sound is often unnatural and includes noticeable clipping artifacts. Moreover, time-domain TSM only preserves the most prominent periodicity. For audio with a wide range of frequency compositions, such as pop music, symphony, and orchestra, less prominent sounds are often discarded in the process.

\subsection{Spectral-domain approach}
Another approach processes the audio within the spectral domain using the short-time Fourier transform (STFT) to convert the frequency information contained in the raw waveform into a more semantic representation in complex numbers \cite{lar99}. Further, the magnitude and phase components can be derived. Unfortunately, unlike the magnitudes, which provide constructive and straightforward audio features, the phases are relatively complex and challenging to model. Moreover, due to the heavy correlation between each phase bin, an additional phase vocoder \cite{fla66} is required to estimate phases and the instantaneous frequencies after carefully relocating STFT bins to prevent specific artifacts known as "phasiness." Although some refined methods \cite{kra12,moi11,nag09} enhance both the vertical and horizontal coherence of the phase, the spectral representation is not inherently scalable, and the iterative phase propagation process in the phase vocoder poses a significant computational overhead.

\subsection{Neural vocoder}
Modeling audio is not a trivial task for neural networks. To illustrate this, let's consider image generation as an example. DCGAN \cite{rad16} is a generative neural network used for synthesizing realistic images with dimensions of $3\times 64\times 64$ pixels. It needs to generate a total of 12,288 pixels. On the other hand, a stereo audio clip lasting 5 seconds, sampled at a rate of 22050 Hz, consists of 220,500 samples. Furthermore, while each pixel in an image is stored in 8 bits, each audio sample is stored in 16 bits, providing 256 times more possible values than an image pixel. Another option to reduce complexity is to decrease the sampling rate. However, according to the Nyquist-Shannon sampling theorem, a low sampling rate leads to significant aliasing.

\begin{figure*}[t]
  \centering
  \vspace{1cm}
  \includegraphics[width=0.95\linewidth]{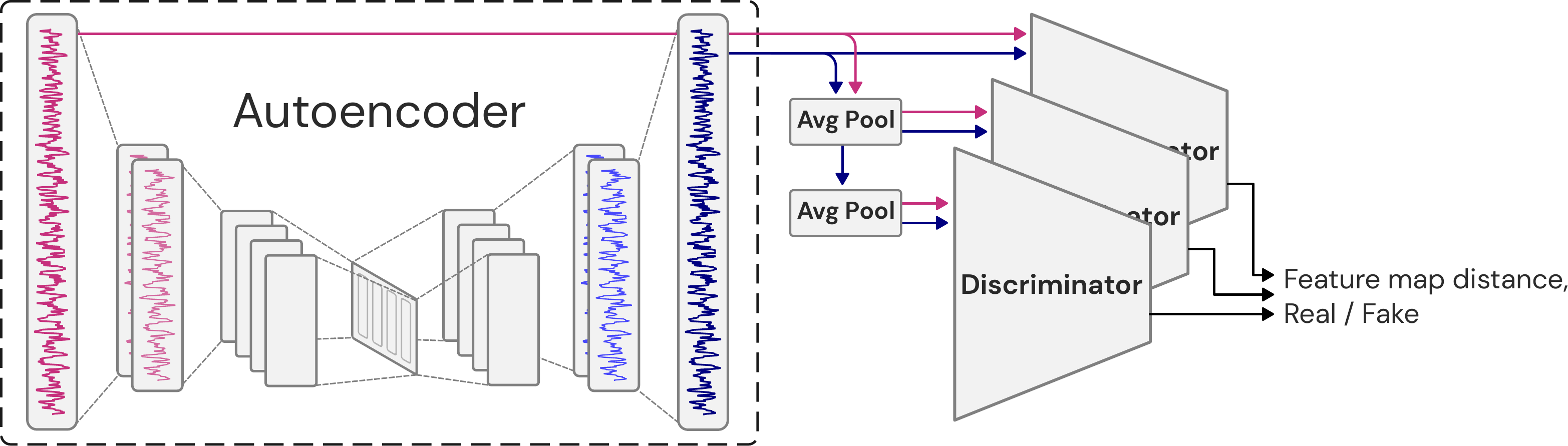}
  \vspace{0.5cm}
  \caption{Overall architecture for the proposed training strategy. The number of channels increases with each layer of convolutional neural networks in the encoder. Both of the input and output of the autoencoder are one-dimensional audio signals. They are consumed by multi-scale discriminators to produce binary predictions. The feature maps of the discriminators are also used during the training to calculate the distance between input and reconstructed audio signal in discriminators' latent space.
  }
  \label{fig:arch}
\end{figure*}

Models that directly generate raw audio waveforms are referred to as vocoders. A vocoder can utilize high-level abstract features, such as linguistic features or spectrograms, for conditioning. The spectrogram represents the magnitude component obtained from the output of STFT. It is easy to model due to its smooth variations in frequency composition over time. As mentioned in Section \ref{sec:td-tsm}, the phases are relatively hard to estimate. Therefore, in applications such as text-to-speech (TTS) pipelines \cite{xu21}, the network often generates the speech spectrogram from given texts and then utilizes a vocoder to synthesize the corresponding audio waveform. Early vocoders, such as Griffin-Lim \cite{gri84} and WORLD \cite{mor16}, were developed.

\subsubsection{Autoregressive and flow-based neural vocoder}
The pioneers of modern neural-based vocoders include WaveNet \cite{aar16}, which predicts the distribution for each audio sample conditioned on all previous ones. However, the autoregressive model runs too slowly to be applied to real-time applications. FloWaveNet \cite{sun18} and WaveGlow \cite{pre19} are neural vocoders that are based on bipartite transforms. They present a faster inference speed and high-quality synthetic audio but require larger models and more parameters to be as expressive as the autoregressive models, thus making them harder to train.

\subsubsection{GAN-based neural vocoder}
WaveGAN \cite{chr18}, MelGAN \cite{kun19}, and VocGAN \cite{yan20} employ the generative adversarial network \cite{goo14} training architecture in which the discriminators are used to measure the divergence of synthetic audio and the real audio and help the generator network synthesize audio samples as realistic as possible. The discriminators usually work at multiple scales to handle different frequency bands in the audio data. This kind of approach allows smaller models to generate high-fidelity audio samples.

%% file: sections/3-method.tex
\label{sec:method}
In this section, we provide a comprehensive discussion of Neuralgram, a novel representation of audio signals, and how it can be used in the TSM. Additionally, we present a brief comparison between Neuralgram and other common representations such as the spectrogram. Finally, we introduce the proposed TSM-Net architecture, which consists of an autoencoder and multi-scale discriminators. The architecture is depicted in Figure \ref{fig:arch}.

\subsection{Latent representation}
We propose a new representation for audio called Neuralgram to provide a novel approach to TSM. The Neuralgram is a temporally compressed feature map extracted from the middle of a neural autoencoder. A Neuralgram is applicable to TSM only when the following conditions are met:
\begin{enumerate}
  \item{An encoder-decoder pair that faithfully reconstructs the audio waveform.}
  \item{A compression ratio that is high enough to put an entire sinusoid of the lowest frequency perceivable into a single sample point in the Neuralgram.}
\end{enumerate}
Instead of directly scaling the raw waveform, which leads to pitch shifting of the audio, we follow a process where we encode the raw waveform as a real-valued Neuralgram, then scale the Neuralgram accordingly. Finally, we decode the scaled Neuralgram to obtain temporally stretched audio while preserving the original pitch. The process is illustrated in Figure \ref{fig:tsm-illustration}. Formally speaking, given an input audio $x$, we obtain the time-scaled signal $\hat{x}$ by
\begin{equation}
  \hat{x} = \mathcal{A}_D(S(\mathcal{A}_E(x), r)),
\end{equation}
where we decompose an optimized autoencoder $\mathcal{A}$ into the encoder $\mathcal{A}_E$ and the decoder $\mathcal{A}_D$, and a scaling function $S$ configured by a factor $r$ is applied to the Neuralgram produced by $\mathcal{A}_E$. In our context, $S$ is a basic cubic interpolation for simplicity. However, $S$ can also be expanded to a more advanced neural technique for super-resolution. Because each sample in the Neuralgram encodes more than an entire sinusoid for each frequency component, resizing the Neuralgram can replicate entire sinusoids in the reconstructed waveform instead of altering their wavelengths and frequencies.

\subsubsection{Neuralgram vs. Spectrogram}
In the literature, most of the works related to the neural vocoder use the spectrogram family as prior conditions. Why should we consider adopting a new representation? These traditional representations encode various frequency information into the same number of sample points in the latent space. This results in different upsample scaling ratios during the decoding process, as each frequency has its own wavelength. The transformation function with variable upsampling ratios is harder to approximate with convolutional generative models. On the contrary, our Neuralgram encodes different frequency components proportionally, which is intuitive for convolutional networks, making the model easier to train.

\input{assets/figures/tsm-illustration}

\subsection{The TSM-Net model}
\label{sec:tsm-method}
\subsubsection{Architecture}
Our model is adapted from the MelGAN \cite{kun19}. In the original generator, the input is a mel-spectrogram, which would be upsampled 256$\times$ to a raw audio waveform. In the proposed model, we increase the upsampling rate to 1024$\times$ to capture the entire sinusoid within a single sample point. This is because in audio with a sampling rate of 22050Hz, the lowest frequency perceived by human ears, a 20Hz sinusoid occupies 1102.5 sample points. The upsampling process involves five stages of upsampling blocks: 8$\times$, 8$\times$, 4$\times$, 2$\times$, and 2$\times$.

Additionally, we prepend a mirror of the modified generator to the front of the generator to obtain the full autoencoder model $A$. Both the encoder $A_E$ and the decoder $\mathcal{A}_D$ are initiated by an aggregating convolutional layer, followed by a deries of downsampling/upsampling stages. Each stage is composed of a dilated downsampling/upsampling convolutional layer and a residual block that also contains a dilated convolutional block and a skip-connection. Formally, the temporal dimensionality of the input signal $x$ satisfy
\begin{equation}
\dim_T(x) = \dim_T(\mathcal{A}_D(\mathcal{A}_E(x))) \approx \frac{SR}{20} \dim_T(\mathcal{A}_E(x)),
\end{equation}
where SR is the sampling rate, which is set to 22050Hz throughout this paper. As discussed in the MelGAN paper, we also use kernel size as a multiple of stride to avoid checkerboard artifacts \cite{ode16}, and weight normalization \cite{lee16} is also used after each layer to improve the sample quality.

\subsubsection{Adversarial losses}
In order to improve the high-frequency fidelity, we employ the same discriminators as the ones in the MelGAN. Three discriminators $(\mathcal{D}_1, \mathcal{D}_2, \mathcal{D}_3)$ operate at different scales simultaneously. With the exception of $\mathcal{D}_1$, which operates at the original scale, downsampling is performed beforehand using stridden average pooling with a kernel size of 4. This arrangement allows each discriminator to more effectively learn features for different frequency ranges of the audio. Next, we formulate the adversarial losses for training the autoencoder $\mathcal{A}$, following the MelGAN approach and using the hinge loss version of the GAN objective \cite{lim17} to penalize only the unstable data distribution.

In order to provide additional information to the autoencoder for utilizing the input condition and prevent mode collapse, we also incorporate the feature-matching loss $\mathcal{L}_{FM}$. This loss minimizes the L1 norm between the discriminator's feature maps of the input and the reconstructed audio. Through empirical observation, we found that minimizing the distance between raw waveforms produces audible artifacts. During the training of the autoencoder, we accumulate the feature matching losses at each intermediate layer of all discriminators, such as
\begin{equation}
  \mathcal{L}_{FM}\left( \mathcal{A}, \mathcal{D}_k\right) =
  \mathbb{E}_x\left[\sum_{i=1}^T \frac{1}{N_i}
  \lVert \mathcal{D}_k^{(i)}(x) - \mathcal{D}_k^{(i)}(\mathcal{A}(x))\rVert_1
  \right],
\end{equation}
where $\mathcal{D}_k^{(i)}$ represents the $i$th layer's feature map output of the $k$th discriminator. $N_i$ is a normalization factor and denotes the number of units in each layer. $T$ represents the number of layers in each discriminator. Our final training objective is given by
\begin{equation}
\min_{\mathcal{D}_k}\sum_{k=1}^3\mathbb{E}_x\left[
  \min\left( 0, 1 - \mathcal{D}_k\left( x\right)\right) +
  \min\left( 0, 1 + \mathcal{D}_k\left( \mathcal{A}\left( x\right)\right)\right)
\right],
\end{equation}
\begin{equation}
\min_{\mathcal{A}}\left(
  \mathbb{E}_x\left[-\sum_{k=1}^3\mathcal{D}_k\left( \mathcal{A}\left( x \right)\right)\right] +
  \lambda\sum_{k=1}^3\mathcal{L}_{FM}\left( \mathcal{A}, \mathcal{D}_k\right)
\right),
\end{equation}
where $\lambda$ controls the strength of $\mathcal{L}_{LM}$ and is set to 10 by default. According to \cite{iso17, mat16}, the noise input is not necessary when the conditioning information is very strong in the generative model.

%% file: assets/figures/tsm-illustration.tex
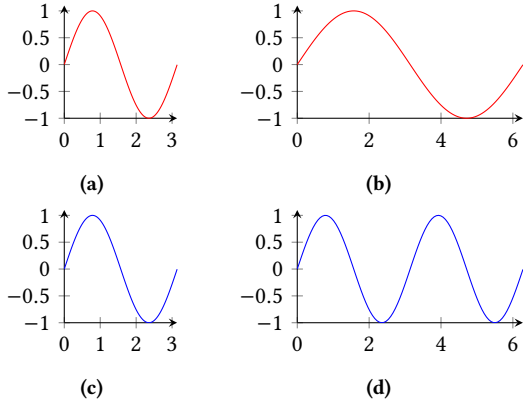
\begin{figure}[t]
\centering
\begin{subfigure}{.3\linewidth}
  \centering
  \begin{tikzpicture}
  \begin{axis}[
    domain=0:3.14,
    axis lines = left,
    legend pos=outer north east,
    width=1.5cm,
    height=1.5cm,
    ymax=1.1,
    scale only axis,
  ]
  \addplot [
    samples=100, 
    color=red,
  ]
  {sin(deg(2*x))};

  \end{axis}
  \end{tikzpicture}
  \caption{}
  \label{fig:sfig1}
\end{subfigure}%
\begin{subfigure}{.6\linewidth}
  \centering
  \begin{tikzpicture}
  \begin{axis}[
    domain=0:6.28,
    axis lines = left,
    legend pos=outer north east,
    width=3cm,
    height=1.5cm,
    ymax=1.1,
    scale only axis,
  ]
  \addplot [
    samples=100, 
    color=red,
  ]
  {sin(deg(x))};

  \end{axis}
  \end{tikzpicture}
  \caption{}
  \label{fig:sfig2}
\end{subfigure}
\begin{subfigure}{.3\linewidth}
  \centering
  \begin{tikzpicture}
  \begin{axis}[
    domain=0:3.14,
    axis lines = left,
    legend pos=outer north east,
    width=1.5cm,
    height=1.5cm,
    ymax=1.1,
    scale only axis,
  ]
  \addplot [
    samples=100, 
    color=blue,
  ]
  {sin(deg(2*x))};

  \end{axis}
  \end{tikzpicture}
  \caption{}
  \label{fig:sfig3}
\end{subfigure}%
\begin{subfigure}{.6\linewidth}
  \centering
  \begin{tikzpicture}
  \begin{axis}[
    domain=0:6.28,
    axis lines = left,
    legend pos=outer north east,
    width=3cm,
    height=1.5cm,
    ymax=1.1,
    scale only axis,
  ]
  \addplot [
    samples=100, 
    color=blue,
  ]
  {sin(deg(2*x))};

  \end{axis}
  \end{tikzpicture}
  \caption{}
  \label{fig:sfig4}
\end{subfigure}
\caption{An illustration for the desire TSM. While doubling the audio duration, the frequency should be kept intact. The original signal contains the sinusoid for a single frequency \ref{fig:sfig1}, \ref{fig:sfig3}.The erroneous signal \ref{fig:sfig2} is the result of directly scaling on the raw waveform, which changes the wavelength of the sinusoid and produces pitch shifting. The desired behavior \ref{fig:sfig4} can be achieved by scaling on the Neuralgram, which compresses the entire sinusoid into one vector. We then repeat the vector and produce two waves through the decoder.}
\label{fig:tsm-illustration}
\end{figure}

%% file: sections/4-exp.tex
We extensively test our model on pop music, classical music, and speech data. We present comprehensive statistics of the feedback collected from our user study, showing that despite its simplicity, the proposed method demonstrates comparable or superior performance on various kinds of audio data.

\subsection{Dataset}
Since our framework does not require any human-annotated data, our model can be trained on any audio dataset. We consider three datasets with various audio contents in our experiment. FMA \cite{kir16} contains a total length of 343 days of Creative Commons-licensed audio, arranged in a hierarchical taxonomy of 161 genres of pop music. Musicnet \cite{joh18} is a collection of 330 freely-licensed classical music recordings. CSTR VCTK Corpus \cite{yam19} includes speech data uttered by 110 English speakers with various accents. All of the audio is resampled to 22050Hz.

\subsection{Implementation details}
We train the model on a single Nvidia Tesla P100 for a week with each dataset. When training on a more complex dataset, such as the FMA dataset, which contains a large amount of Pop music, the wide frequency range in Pop music makes training the GAN from scratch more difficult. We pre-train the autoencoder on the classical music dataset, Musicnet, to stabilize the training progress. In contrast, a pre-trained discriminator cannot effectively guide the autoencoder in performing a proper reconstruction. We attach the source code\footnote{\url{https://github.com/tsmnet-mmasia23/tsmnet}.} accompanying this paper to encourage reproducibility and enhancements.

\input{assets/tables/mos}

\subsection{Time-scaled modification}
\label{sec:user-study}
\subsubsection{Setup}
To effectively evaluate the quality of the stretched audio, we employ a Monte Carlo approach to collect user feedback on the test samples. We randomly select twenty 10-second (at most) samples from the FMA, Musicnet, and VCTK datasets for the listening test. In each round of the test, 10 out of the total of 60 test samples are drawn to be presented to the participants. We varied the speed of the audio using a factor $r$, randomly chosen from intervals $[0.5, 0.95]$ with an interval of $0.05$ and $[1.1, 2.0]$ with an interval of $0.1$. To evaluate the performance of the generated audio, we utilized the mean opinion score (MOS) and recruited 68 participants with diverse backgrounds, who contributed ratings for a total of 580 audio samples. For each audio sample, participants were presented with the original audio and five audio samples stretched to the same speed using various methods. They were asked to rate the generated audio on a scale of 1 (poor) to 5 (excellent) in terms of pitch correctness and overall audio quality. The listening test can be experienced on our website\footnote{\url{https://tsmnet-mmasia23.web.app}.}

\subsubsection{Results}
Table \ref{tab:mos} indicates that despite the simplicity of our method, the participants consider the samples generated by our method comparable to the state-of-the-art approach on both musical datasets and speech datasets. Furthermore, Figure \ref{fig:hist} shows the histogram of MOS ratings for the audio samples stretched by the proposed method. The collected data roughly follow a normal distribution, indicating the reliability of our subjective test.

\input{assets/figures/hist}

\subsection{Study on different compression ratios}
As mentioned in Section \ref{sec:tsm-method}, a compression ratio (CR) of 1024$\times$ is large enough for the lowest perceivable frequency. We would like to investigate whether a lower CR results in pitch-shifting in TSM. We examine different models with CRs of 256$\times$, 512$\times$, and the original 1024$\times$. Intuitively, a lower CR should result in smaller reconstruction errors due to reduced information loss. However, the pitch-shifting effect occurs across a wider range of frequency bands when the CR is small. The pitch-shifting effect gradually emerges at lower frequencies and seldom occurs in the high frequencies. We recommend listening to the audio samples on our website\footnote{\url{https://tsmnet-mmasia23.github.io}.} to understand this interesting phenomenon.

In addition to the qualitative evaluation, we also conduct a listening test as set up in Section \ref{sec:user-study}, in which we compare the audio samples stretched by the three variants of TSM-Net. The results in Table \ref{tab:cr} show that sufficiently high compression ratios are crucial to prevent pitch-shifting and ensure acceptable audio quality. Figure \ref{fig:mos-rate} also indicates that the $1024\times$ CR setting makes TSM-Net stretch audio to various speeds with better quality preservation. The figure also shows that our models perform worse when stretching the audio toward more extreme speeds. We hypothesize that the problem mainly arises from an overly naive interpolation algorithm. We leave the investigation of this issue for future work.

\input{assets/tables/cr}

\input{assets/figures/mos-speed}

\subsection{Inference time}
As a neural network-based approach, our method can easily leverage the highly parallel GPU computation unit. To study the processing speed improvement, we record the inference time for generating 1200 audio samples used in the listening test. We report the numbers in milliseconds per second of audio signal for the baseline methods and the variants of TSM-Net. As shown in Table \ref{tab:time}, our method reduces the computation time by a large margin compared to the baseline methods, as it does not rely on time-consuming phase alignment or phase propagation, and it can leverage the power of GPU.

\input{assets/tables/time}

\subsection{Additional training techniques}

We also use two additional metrics to monitor our training but they are not included in the training losses.
\begin{enumerate}
  \item{Audio Reconstruction (AR). AR is the L1 norm between the real and reconstructed raw audio waveform.}
  \item{Neuralgram Reconstruction (NR). NR is the L1 norm between the Neuralgrams encoded from the real and reconstructed raw audio waveform, i.e., the reconstructed Neuralgram has 1.5 passes through the autoencoder.}
\end{enumerate}

We note that the model is not guaranteed to converge in each run. The ideal loss for the discriminator tends to fluctuate around 6. If the discriminator's loss decreases drastically, both the autoencoder's loss and the feature matching loss fail to return to a healthy value, resulting in rapid divergence and poor quality. We can verify this phenomenon by examining AR and NR. Both reconstruction losses increase after the discriminator gains dominance in the training process. Notably, we can perceive background noises in the reconstructed audio samples.

%% file: assets/tables/mos.tex
\begin{table}
\centering
\caption{Mean opinion score on three datasets.} 
\begin{tabular}{l c c c | c}\toprule
Method & FMA & Musicnet & VCTK & Average \\
\midrule
WSOLA \cite{ver93} & 2.8 & 3.31 & 3.12 & 3.08\\
PV-TSM \cite{moi11} & 2.89 & 3.24 & 3.09 & 3.07\\
TSM-Net (ours) & 2.87 & 3.20 & 3.12 & 3.06 \\\bottomrule
\end{tabular}
\label{tab:mos}
\end{table}

%% file: assets/figures/hist.tex
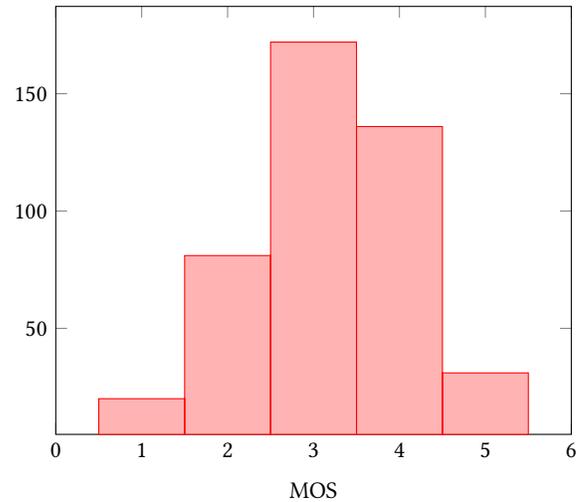
\begin{figure}
\centering
\begin{tikzpicture}
\pgfplotstableread[col sep=comma]{assets/data/hist.csv}\datatable
\begin{axis}[
  xlabel={MOS}
]
\addplot+[
  hist={bins=5, data min=0.5, data max=5.5, density=false},
  mark=none,
  color=red,
  fill=red!30,
] table[y=mos] {\datatable};
\end{axis}
\end{tikzpicture}
\caption{The MOS histogram for TSM-Net.}
\label{fig:hist}
\end{figure}

%% file: assets/tables/cr.tex
\begin{table}
\centering
\caption{Mean opinion score on three datasets. $256\times$ and $512\times$ indicate an architecture with a compressing ratio of 256 and 512, respectively. The default compressing ratio of the proposed method is $1024\times$.} 
\begin{tabular}{l c c c | c}
\toprule
Compression ratio & FMA & Musicnet & VCTK & Average \\
\midrule
$256\times$ & 2.78 & 3.19 & 2.98 & 2.98 \\
$512\times$ & 2.78 & 3.14 & 3.01 & 2.97 \\
$1024\times$ (ours) & 2.87 & 3.20 & 3.12 & 3.06\\
\bottomrule
\end{tabular}
\label{tab:cr}
\end{table}

%% file: assets/figures/mos-speed.tex
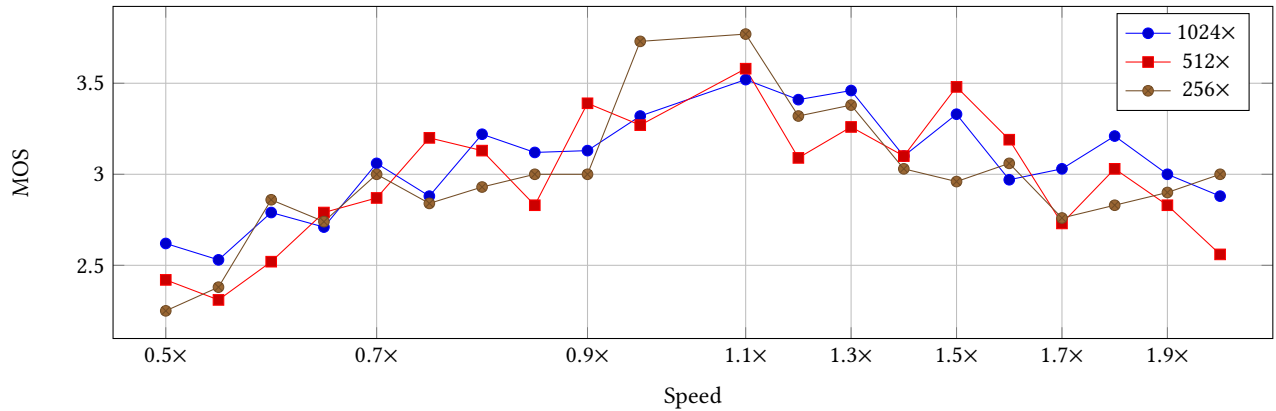
\begin{figure*}
\centering
\begin{tikzpicture}
\pgfplotsset{
    x coord trafo/.code={
        \pgfmathparse{#1>=1 ? #1+1 : #1*2}
    },
    x coord inv trafo/.code={
        \pgfmathparse{#1>=2 ? #1-1 : #1/2}
    }
}
\pgfplotstableread[col sep=comma]{assets/data/mos-speed.csv}\datatable
\begin{axis}[
  xlabel={Speed},
  ylabel={MOS},
  xmin=0.45, xmax=2.1, 
  grid=major,
  xtick={0.5,0.7,0.9,1.1,1.3,1.5,1.7,1.9},
  xticklabel={$\pgfmathprintnumber{\tick}\times$},
  width=17cm, 
  height=6cm 
]
\addplot table[x={speed}, y={1024x}] {\datatable};
\addlegendentry{$1024\times$}
\addplot table[x={speed}, y={512x}] {\datatable};
\addlegendentry{$512\times$}
\addplot table[x={speed}, y={256x}] {\datatable};
\addlegendentry{$256\times$}
\end{axis}
\end{tikzpicture}
\caption{Average MOS on each speed. $256\times$ and $512\times$ indicate an architecture with a compressing ratio of 256 and 512, respectively. The default compressing ratio of the proposed method is $1024\times$.}
\label{fig:mos-rate}
\end{figure*}

%% file: assets/tables/time.tex
\begin{table}
\centering
\caption{Average inference time per 1-sec audio generation.}
\begin{tabular}{l c}
\toprule
Method & Average inference time \\
\midrule
WSOLA \cite{ver93} &                  16.4807 ms/sec \\
PV-TSM \cite{moi11} & 15.4474 ms/sec \\
TSM-Net $1024\times$ & 1.8954 ms/sec \\
TSM-Net $512\times$ & 1.7550 ms/sec \\
TSM-Net $256\times$ & 1.5599 ms/sec \\
\bottomrule
\end{tabular}
\label{tab:time}
\end{table}

%% file: sections/5-conclusion.tex
In this paper, we introduce a custom neural network model and a novel audio representation for the time-scale modification (TSM). The proposed method demonstrates a simple yet efficient approach to manipulating audio contents temporally using the power of the neural compressor. Our method mitigates or solves the issues found in the traditional TSM, such as the harmonic alignment problem, the background sound loss, and the phasiness. However, our method sometimes produces other artifacts, which make the human evaluators consider the stretched audio less preferred. We attribute this issue to the insufficient model capacity and the naive choice of interpolation, which are practical limitations under our theoretical proposition. To improve audio quality, our method can be further incorporated with advancements in other domains, such as neural interpolation.

While the research of TSM had been silent for a long time, this ubiquitous technology is now used in our everyday life. We believe our work opens new possibilities for state-of-the-art TSM algorithms, allowing for further advancements and applications in this field.